\documentclass[12pt]{article}
\usepackage{epsfig}

\begin{document}

\begin{center}

\vspace*{1.0cm}

{\Large \bf{Search for $2\beta$ decay of $^{106}$Cd with enriched
$^{106}$CdWO$_4$ crystal scintillator in coincidence with four
HPGe detectors}}

\vskip 1.0cm

{\bf

P.~Belli$^{a}$, R.~Bernabei$^{a,b,}$\footnote{Corresponding
author.
                    {\it E-mail address:} rita.bernabei@roma2.infn.it.},
V.B.~Brudanin$^{c}$,
F.~Cappella$^{d}$,
V.~Caracciolo$^{d}$,
R.~Cerulli$^{d}$,
D.M.~Chernyak$^{e}$,
F.A.~Danevich$^{e}$,
 S.~d'Angelo$^{a,b,}$\footnote{Deceased},
 A.~Di~Marco$^{a,b}$,
 A.~Incicchitti$^{f,g}$,
M.~Laubenstein$^{d}$,
V.M.~Mokina$^{e}$,
D.V.~Poda$^{e,h}$,
O.G.~Polischuk$^{e,f}$,
V.I.~Tretyak$^{e,f}$,
I.A.~Tupitsyna$^{i}$

}

\vskip 0.3cm

$^{a}${\it INFN, sezione di Roma ``Tor Vergata'', I-00133 Rome,
Italy}

$^{b}${\it Dipartimento di Fisica, Universit\`a di Roma ``Tor Vergata'', I-00133 Rome, Italy}

$^{c}${\it Joint Institute for Nuclear Research, 141980 Dubna, Russia}

$^{d}${\it INFN, Laboratori Nazionali del Gran Sasso, I-67100 Assergi (AQ), Italy}

$^{e}${\it Institute for Nuclear Research, MSP 03680 Kyiv, Ukraine}

$^{f}${\it INFN, sezione di Roma, I-00185 Rome, Italy}

$^{g}${\it Dipartimento di Fisica, Universit\`a di Roma ``La Sapienza'', I-00185 Rome, Italy}

$^{h}${\it Centre de Sciences Nucl\'eaires et de Sciences de la Mati\`ere, 91405 Orsay, France}

$^{i}${\it Institute of Scintillation Materials, 61001 Kharkiv, Ukraine}

\end{center}

\vskip 0.5cm

\begin{abstract}

A radiopure cadmium tungstate crystal scintillator, enriched in
$^{106}$Cd to 66\%, with mass of 216 g ($^{106}$CdWO$_4$), was
used to search for double beta decay processes in $^{106}$Cd in
coincidence with four ultra-low background high purity germanium
detectors in a single cryostat. New improved limits on the double
beta processes in $^{106}$Cd have been set on the level of
$10^{20}- 10^{21}$ yr after 13085 h of data taking. In particular,
the half-life limit on the two neutrino electron capture with
positron emission, $T_{1/2}^{2\nu\varepsilon\beta^+}\geq 1.1\times
10^{21}$~yr, has reached the region of theoretical predictions.
With this half-life limit the effective nuclear matrix element for
the $2\nu\varepsilon\beta^+$ decay is bounded as
$M^{2\nu\varepsilon\beta^+}_{eff}\le 1.1$. The resonant
neutrinoless double electron captures to the 2718 keV, 2741 keV
and 2748 keV excited states of $^{106}$Pd are restricted at the
level of $T_{1/2} \geq (8.5\times10^{20}-1.4\times10^{21}$) yr.

\end{abstract}

\vskip 0.4cm

\noindent PACS number(s): 29.40.Mc, 23.40.-s

\vskip 0.4cm


\section{INTRODUCTION}

Experiments to search for neutrinoless double beta ($0\nu2\beta$)
decay are considered as a promising way to study properties of
neutrino and weak interactions, test the lepton number
conservation \cite{{Vergados:2012,Barrea:2012,Rodejohann:2012}}.
In addition, the process can be mediated by right-handed currents
in weak interaction, existence of massless (or very light)
Nambu-Goldstone bosons (majorons), and many other effects beyond
the Standard Model
\cite{Vergados:2012,Rodejohann:2012,Deppisch:2012,Bilenky:2015}.

Experimental efforts are concentrated mainly on the double beta
decay with emission of two electrons (see reviews
\cite{Tretyak:1995,Tretyak:2002,Elliott:2012,Giuliani:2012,Saakyan:2013,Cremonesi:2014,Gomes:2015,Sarazin:2015}).
The results of the experiments to search for double beta processes
with decrease of nuclear charge: the capture of two electrons from
atomic shells ($2\varepsilon$), electron capture with positron
emission ($\varepsilon\beta^+$), and emission of two positrons
($2\beta^+$) are substantially more modest (we refer reader to the
reviews \cite{Tretyak:1995,Tretyak:2002,Maalampi:2013} and
references 11--27 in \cite{Belli:2012}). Even the allowed two
neutrino mode of the double beta plus processes is not yet
detected unambiguously: there are only indications on two neutrino
double electron capture in $^{130}$Ba
\cite{Meshik:2001,Pujol:2009} and $^{78}$Kr \cite{Gavrilyuk:2013}
with the half-lives on the order $10^{20}-10^{21}$ yr. At the same
time a strong motivation to search for neutrinoless
$\varepsilon\beta^+$ decay is related with the possibility to
refine the mechanism of the $0\nu2\beta^-$ decay when observed:
whether it appears due to the Majorana mass of neutrino or due to
the contribution of the right-handed admixtures in the weak
interaction \cite{Hirsh:1994}. In addition, experimental data on
the two neutrino decay could be useful to improve theoretical
calculations of the decay probability.

\begin{figure}[htb]
\begin{center}
\mbox{\epsfig{figure=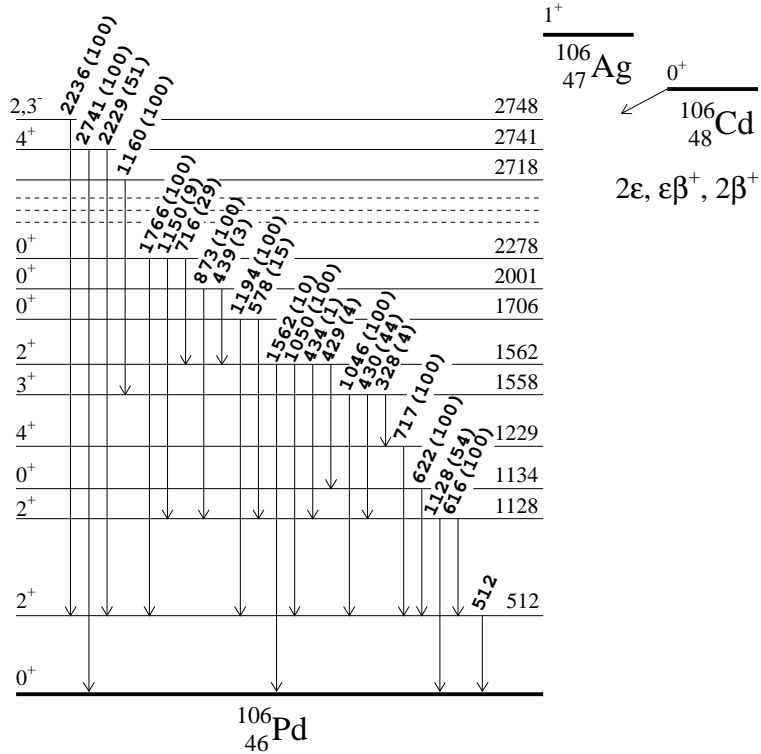,height=10.0cm}} \caption{Simplified
decay scheme of $^{106}$Cd \cite{Frenne:2008} (levels at
$2283-2714$ keV are omitted). Energies of the excited levels and
of the emitted $\gamma$ quanta are in keV. Relative intensities of
$\gamma$ quanta are given in parentheses.} \label{fig01}
\end{center}
\end{figure}

The isotope $^{106}$Cd (energy of decay $Q_{2\beta} = 2775.39(10)$
keV \cite{Wang:2012}, natural isotopic abundance $\delta =
1.25(6)\%$ \cite{Berglund:2011}) is one of the most suitable
nuclei to search for the double beta plus processes. In addition
the isotope $^{106}$Cd is favored to search for resonant
$0\nu2\varepsilon$ transitions to excited levels of $^{106}$Pd
when there is a coincidence between the released energy and the
energy of an excited state \cite{Belli:2012,Krivoruchenko:2011}.
The decay scheme of $^{106}$Cd is presented in Fig. \ref{fig01}.

\nopagebreak

The nuclide $^{106}$Cd is one of the most studied (see a detailed
review of the previous investigations in \cite{Belli:2012}). At
present there are three experiments in progress aiming at search
for double beta decay of $^{106}$Cd. The TGV-2 experiment in the
Modane underground laboratory in France utilizes 32 planar HPGe
detectors with 16 thin foils of enriched $^{106}$Cd
($\delta=75\%$) installed between the detectors. Recently new
enriched foils with a higher isotopic concentration of $^{106}$Cd
$\delta=99.6\%$ were installed. The sensitivity of the experiment
is at the level of $\lim T_{1/2}\sim 10^{20}$ yr
\cite{Rukhadze:2011a,Rukhadze:2011b,Briancon:2015}. The main
advantage of the TGV-2 experiment is a high sensitivity to the
$2\nu2K$ decay of $^{106}$Cd with the theoretically shortest
half-life of the three allowed channels.

The COBRA experiment uses an array of CdZnTe room temperature
semiconductors $\approx 1$ cm$^3$ each at the Gran Sasso
underground laboratory in Italy. The sensitivity of the experiment to
the double beta processes in $^{106}$Cd is at the level of $\lim
T_{1/2}\sim 10^{18}$ yr \cite{Ebert:2013}.

The third experiment, also carried out at the Gran Sasso
laboratory, utilizes a radiopure cadmium tungstate crystal
scintillator with mass 216 g produced from cadmium enriched in
$^{106}$Cd to 66\% ($^{106}$CdWO$_4$) \cite{Belli:2010}. At the
first stage of the experiment the sensitivity at the level of
$\lim T_{1/2}\sim 10^{20}-10^{21}$ yr was reached for different
channels of double beta decay of $^{106}$Cd \cite{Belli:2012}. To
increase the experimental sensitivity to the $2\beta$ processes
with emission of $\gamma$ quanta, the $^{106}$CdWO$_4$
scintillator was placed inside a low background HPGe detector with
four Ge crystals. Here we report the final results of the
experiment with the $^{106}$CdWO$_4$ scintillation detector
operated in coincidence (anticoincidence) with the four crystals
HPGe $\gamma$ detector. Preliminary results of the experiment were
presented in conference proceedings
\cite{Tretyak:2014,Polischuk:2015a,Danevich:2015a,Tretyak:2016}.

\section{THE EXPERIMENT}

\begin{figure}[htb]
\begin{center}
\mbox{\epsfig{figure=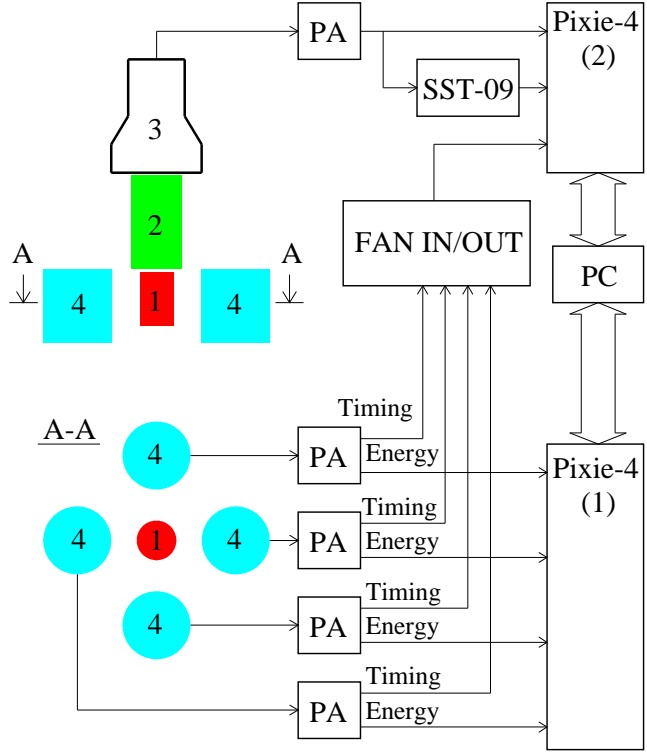,height=10.0cm}} \caption{(Color
online) Low background $^{106}$CdWO$_4$ crystal scintillator (1)
viewed through PbWO$_4$ light-guide (2) by PMT (3). The
scintillator is installed between HPGe detectors (4). Scheme of
the electronic chain: (PA) preamplifiers; (FAN IN/OUT) linear
FAN-IN/FAN-OUT; (SST-09) home-made electronic unit to provide
triggers for cadmium tungstate scintillation signals; (Pixie-4)
four-channel all digital spectrometers; (PC) personal computer.}
\label{fig02}
\end{center}
\end{figure}

The $^{106}$CdWO$_4$ crystal scintillator was viewed through a
lead tungstate (PbWO$_4$) crystal light-guide ($\oslash40 \times
83$ mm) by 3 inches low radioactive photomultiplier tube (PMT)
Hamamatsu R6233MOD (see Fig. \ref{fig02}). The PbWO$_4$ crystal
was developed from deeply purified \cite{Boiko:2011}
archaeological lead \cite{Danevich:2009}. The detector was
installed in an ultra-low background GeMulti HPGe $\gamma$
spectrometer of the STELLA (SubTErranean Low Level Assay) facility
\cite{Laubenstein:2004} at the Gran Sasso underground laboratory
of the INFN (Italy) at the depth of 3600 m of water equivalent.
Four HPGe detectors of the GeMulti set-up are mounted in one
cryostat with a well in the centre. The volumes of the HPGe
detectors are approximately 225 cm$^3$ each.

An event-by-event data acquisition system is based on two
four-channel digital spectrometers (DGF Pixie-4, XIA, LLC). One
device (marked (1) in Fig. \ref{fig02}) is used to provide
spectrometric data for the HPGe detectors, while the second
Pixie-4 (2) acts as a 14-bit waveform digitizer to acquire signals
from the $^{106}$CdWO$_4$ detector at the rate of 18.8 MSPS over a
time window 54.8 $\mu$s. The second Pixie-4 unit records also
trigger signals from the home made unit SST-09, which provides the
triggers only if the signal amplitude in the $^{106}$CdWO$_4$
detector exceeds $\sim0.6$ MeV to avoid acquisition of a large
amount of data caused by the decays of $^{113}$Cd$^m$ ($Q_\beta =
586$ keV) present in the $^{106}$CdWO$_4$ crystal
\cite{Belli:2012,Danevich:2013}. The signals from the timing
outputs of the HPGe detectors after summing are fed to the third
input of the second Pixie-4 digitizer to select coincidence
between the $^{106}$CdWO$_4$ and HPGe detectors off-line.

\nopagebreak

The $^{106}$CdWO$_4$ and HPGe detectors were calibrated with
$^{22}$Na, $^{60}$Co, $^{137}$Cs and $^{228}$Th $\gamma$ sources.
The energy resolution of the $^{106}$CdWO$_4$ detector can be
described by the function: FWHM = $(21.7 \times E_\gamma)^{1/2}$,
where FWHM and $E_\gamma$ are given in keV. The energy resolution
of the HPGe detectors during the experiment was FWHM~$\approx 2-3$
keV for the 1332 keV $\gamma$ quanta of $^{60}$Co.

\begin{figure}[htb]
\begin{center}
\mbox{\epsfig{figure=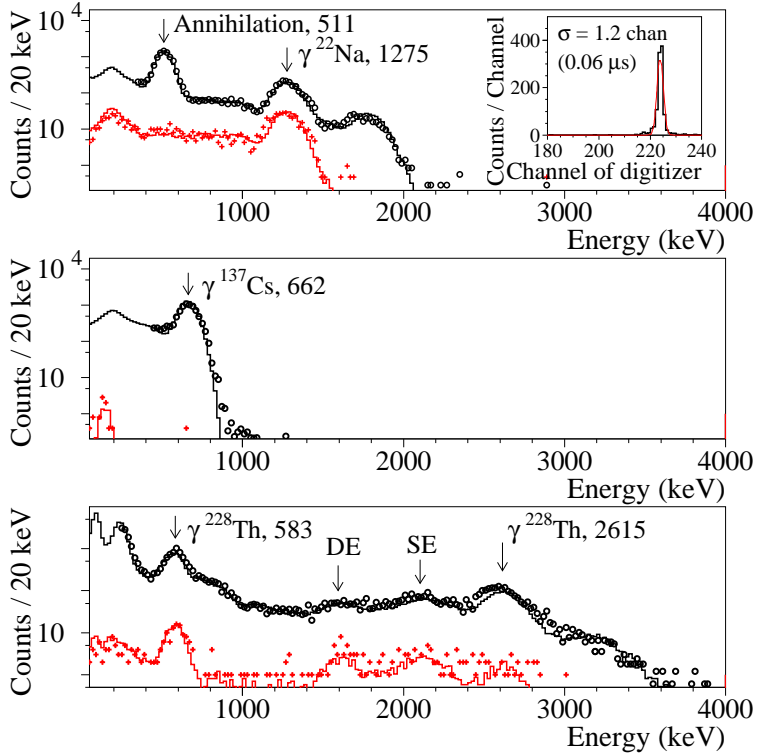,height=10.0cm}} \caption{(Color
online) Energy spectra of $^{22}$Na (upper figure), $^{137}$Cs
(middle figure) and $^{228}$Th (lower figure) $\gamma$ sources
accumulated by the $^{106}$CdWO$_4$ detector: with no coincidence
(black circles), and in coincidence with energy 511 keV in the
HPGe detector (red crosses). The data simulated by using the EGS4
Monte Carlo code are drawn by solid lines. (Inset) Distribution of
the $^{106}$CdWO$_4$ detector pulses start positions relatively to
the HPGe signals with the energy 511 keV accumulated with
$^{22}$Na source (the time shift of the $^{106}$CdWO$_4$ signals
by $\simeq220$ channels is due to the tuning of the digitizer to
provide baseline data).} \label{fig03}
\end{center}
\end{figure}

Energy spectrum and distribution of the start positions of the
$^{106}$CdWO$_4$ detector pulses relatively to the HPGe signals
accumulated with the $^{22}$Na $\gamma$ source (see upper panel in
Fig. \ref{fig03}) demonstrate presence of coincidences between the
$^{106}$CdWO$_4$ and HPGe detectors under the condition that the
energy of events in the HPGe detectors is equal to 511 keV (energy
of annihilation $\gamma$ quanta), while practically there is no
coincidence in the data accumulated with $^{137}$Cs. As one can
see in Fig. \ref{fig03}, the distributions simulated by the EGS4
code \cite{Nelson:1985} are in agreement with the experimental
data obtained with $^{22}$Na, $^{137}$Cs and $^{228}$Th gamma
sources. It should be stressed that the energy threshold of the
$^{106}$CdWO$_4$ detector in the coincidence mode ($\approx50$
keV) is much lower than that in the anticoincidence mode since the
data acquisition in the coincidence mode is triggered by the
signals in the HPGe detectors.

\nopagebreak

\section{RESULTS AND DISCUSSION}

\subsection{Data analysis}

\begin{figure}[htb]
\begin{center}
\mbox{\epsfig{figure=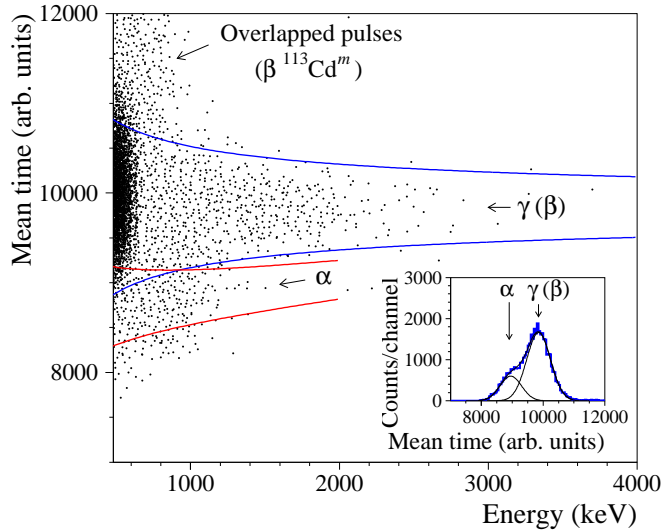,height=7.0cm}} \caption{(Color
online) Mean time (see text) versus the energy accumulated over
571 h with the $^{106}$CdWO$_4$ crystal scintillator in the
low-background set-up. The plus-minus two sigma interval for mean time values
corresponding to gamma quanta (beta particles) and one sigma
interval for alpha particles are depicted. Events with the mean
time values greater than $\approx 1.1\times10^4$ can be explained
by the overlap of events (mainly of beta decays of $^{113}$Cd$^m$
in the crystal). Inset: Distribution of the mean times in the
energy interval $0.9-1.1$ MeV demonstrates the ability of
pulse-shape discrimination between $\gamma$($\beta$) and $\alpha$
events.} \label{fig04}
\end{center}
\end{figure}

The mean-time pulse-shape discrimination method (see, e.g.,
\cite{Bardelli:2008}) was used to discriminate $\gamma(\beta)$
events from $\alpha$ events caused by internal contamination of
the crystal by uranium and thorium. The scatter plot of the mean
time versus energy of the background events accumulated over 571 h
by the $^{106}$CdWO$_4$ detector is depicted in Fig. \ref{fig04}.
The efficiency of the pulse-shape discrimination is worse than
that in the previous experiment \cite{Belli:2012} due to the lower
light collection efficiency with the PbWO$_4$ light-guide.
Nevertheless, the distribution of the mean time for the events
with energies in the range $0.9-1.1$ MeV (see Inset in Fig.
\ref{fig04}) justifies pulse-shape discrimination between $\alpha$
particles and $\gamma$ quanta ($\beta$ particles).  The mean time
for $\gamma$ quanta was measured with the $^{228}$Th $\gamma$ source in the
energy range $0.5-2.6$ MeV as $\tau_{\gamma}=9843$. The energy
dependence of the mean time distribution sigma
($\sigma^{\tau}_{\gamma}$) was determined as
$\sigma^{\tau}_{\gamma}=3376/\sqrt{0.1\times E_{\gamma}}$, where
$E_{\gamma}$ is energy of gamma quanta in keV. The pulse-shape
discrimination was also used to reject overlapped pulses (mainly
caused by pileups of the $^{113}$Cd$^m$ $\beta$ decay events) and
pileups of the PbWO$_4$ scintillation signals (characterized by
rather short decay time at the level of $\approx 0.3~\mu$s
\cite{Bardelli:2008}) with the $^{106}$CdWO$_4$ signals (effective
average decay time $\approx 13~\mu$s \cite{Bardelli:2006}).

\nopagebreak

\begin{figure}[htb]
\begin{center}
\mbox{\epsfig{figure=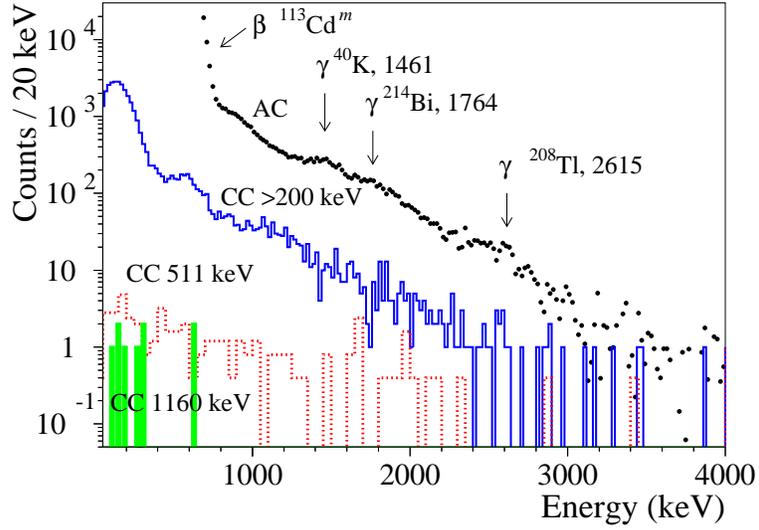,height=7.0cm}} \caption{(Color
online) Energy spectra of $\gamma$ and $\beta$ events accumulated
over 13085 h by the $^{106}$CdWO$_4$ detector in anticoincidence
with the HPGe detectors (``AC"), in coincidence with event(s) in
at least one of the HPGe detectors with the energy $E >200$ keV
(``CC $>$200 keV"), $E=511\pm 3\sigma_{511}$ keV (``CC 511 keV"),
and $E=1160\pm 2.3\sigma_{1160}$ keV (``CC 1160 keV").}
\label{fig05}
\end{center}
\end{figure}

The energy spectra accumulated over 13085 h by the
$^{106}$CdWO$_4$ detector in anticoincidence with the HPGe
detectors, in coincidence with event(s) in at least one of the
HPGe detectors with energy $E > 200$ keV, $E=511$ keV and $E=1160$
keV are presented in Fig. \ref{fig05}. The events in the
anticoincidence spectrum were selected by using the following
cuts: 1) there is no signal(s) in the HPGe detectors, 2) the mean
time ($\tau$) value of a scintillation pulse is within the plus-minus two
sigma interval around the central value:
$\tau_{\gamma}-2\sigma^{\tau}_{\gamma}<\tau<\tau_{\gamma}+2\sigma^{\tau}_{\gamma}$ (the interval is shown in Fig.
\ref{fig04}). The pulse-shape discrimination cut selects 95.5\% of
$\gamma(\beta)$ events. The spectrum of the $^{106}$CdWO$_4$
detector in coincidence with the energy release in at least one of
the HPGe detectors of more than 200 keV was built with the help of
the same pulse-shape discrimination cut. The energy spectra of the
$^{106}$CdWO$_4$ detector in coincidence with the signal(s) in the
HPGe detectors with energy 511 keV (1160 keV) were built by using
the following cuts: 1) there is event(s) in at least one of the
HPGe detectors with energy $E=511~\pm~3\sigma_{511}$ keV
($E=1160~\pm~2.3\sigma_{1160}$ keV) where $\sigma_{511}$ and
$\sigma_{1160}$ are the energy resolutions of the HPGe detectors
for the annihilation peak and for gamma quanta with energy 1160
keV, respectively; 2) the signals in the $^{106}$CdWO$_4$ and HPGe
detectors coincide within the plus-minus three sigma time interval (see Inset of
Fig. \ref{fig03}, the time cut selects 99.7\% of the coinciding
events); 3) the mean time value of a $^{106}$CdWO$_4$
scintillation pulse is within the $\pm 3\sigma^{\tau}_{\gamma}$ interval. The pulse-shape discrimination cut selects 99.7\% of the
$\gamma(\beta)$ events.

The data accumulated by the $^{106}$CdWO$_4$ detector in
anticoincidence with the HPGe detectors confirmed the assumption
about surface contamination of the $^{106}$CdWO$_4$ crystal by
$^{207}$Bi \cite{Belli:2012}. The $\gamma$ peaks of $^{207}$Bi
observed in \cite{Belli:2012} disappeared after the cleaning of
the scintillator by potassium free detergent and ultra-pure nitric
acid.

\nopagebreak

\begin{figure}[htb]
\begin{center}
\mbox{\epsfig{figure=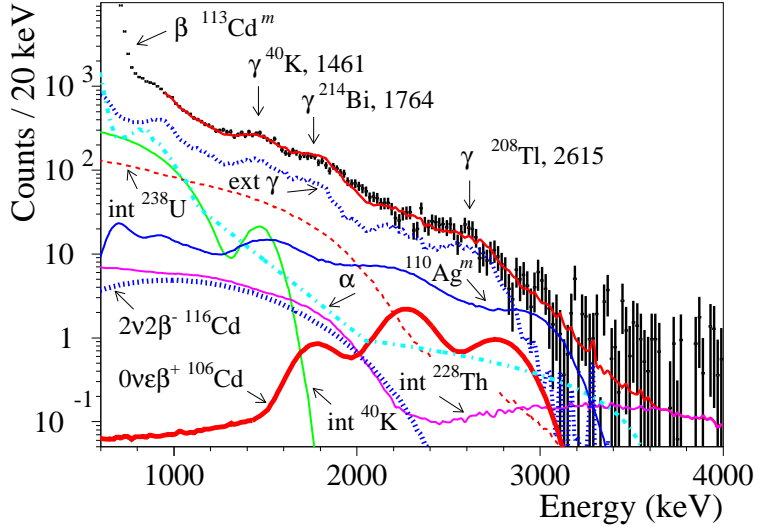,height=7.0cm}} \caption{(Color
online) The energy spectrum of the $\gamma$ and $\beta$ events
accumulated over 13085 h in the low background set-up with the
$^{106}$CdWO$_4$ crystal scintillator (points) together with the
background model (red continuous superimposed line). The main
components of the background are shown: the distributions of
internal and external ("ext $\gamma$") $^{40}$K, $^{232}$Th and
$^{238}$U, distribution of residual alpha particles ($\alpha$), cosmogenic $^{110}$Ag$^m$ in
the $^{106}$CdWO$_4$ scintillator (with activity 0.3 mBq/kg) and $2\nu2\beta^{-}$ decay of $^{116}$Cd. The excluded
distribution of the $0\nu\varepsilon\beta^+$ decay of $^{106}$Cd
to the ground state of $^{106}$Pd with the half-life
$T_{1/2}=1.5\times10^{21}$ yr is shown too.} \label{fig06}
\end{center}
\end{figure}

The anticoincidence spectrum was fitted in the energy interval
$940-3980$ keV ($\chi^2$/n.d.f.$=157/118=1.33$, where n.d.f. is
the number of degrees of freedom) by the model built from the
energy distributions simulated by EGS4 code \cite{Nelson:1985}.
The model includes radioactive contamination of the
$^{106}$CdWO$_4$ crystal scintillator
\cite{Belli:2012,Danevich:2013}, external gamma quanta from the
materials of the set-up ($^{40}$K, $^{232}$Th, $^{238}$U in the
cryostat of the HPGe detector, PMT, PbWO$_4$ light-guide,
$^{26}$Al in the aluminium well of the cryostat), distribution of
alpha particles (which passed the pulse-shape discrimination cut
to select beta and gamma events), cosmogenic $^{110}$Ag$^m$ in
the $^{106}$CdWO$_4$ scintillator and two neutrino $2\beta^{-}$ decay
 of $^{116}$Cd with the half-life $T_{1/2}=2.62\times10^{19}$ yr \cite{Danevich:2016} present in the $^{106}$CdWO$_4$ crystal
with the isotopic abundance $\delta = 1.5\%$ \cite{Belli:2010}.
The result of the fit and the main components of the background
are shown in Fig. \ref{fig06}.

\nopagebreak

\begin{figure}[htb]
\begin{center}
\mbox{\epsfig{figure=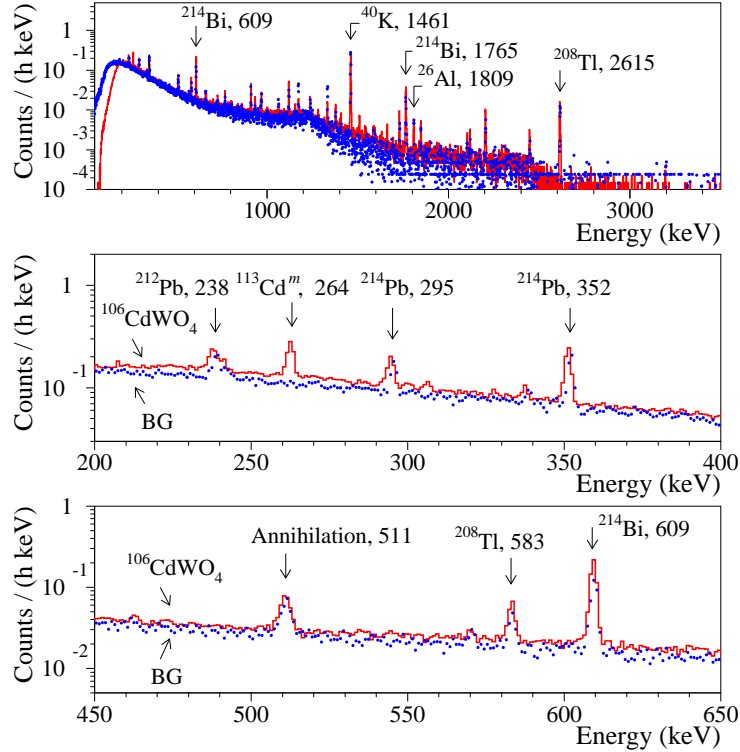,height=10.0cm}} \caption{(Color
online) Energy spectrum accumulated over 13085 h by the low
background HPGe $\gamma$ detector with the $^{106}$CdWO$_4$
scintillation detector inside (solid red histogram) and background
data measured over 4102 h (blue dots). The energy spectra in the
$200-400$ keV (middle part) and $450-650$ keV (lower part) energy
intervals. Energies of $\gamma$ quanta are in keV.} \label{fig07}
\end{center}
\end{figure}

The energy spectrum accumulated by the HPGe detector is presented
in Fig. \ref{fig07} together with the background data taken over
4102 h. The counting rate of the HPGe detector with the
$^{106}$CdWO$_4$ detector inside exceeds slightly the background
counting rate. Some excess (at the level of 30\% -- 170\%
depending on the energy of $\gamma$ quanta) is observed in the
peaks of $^{214}$Bi and $^{214}$Pb (daughters of $^{226}$Ra from
the $^{238}$U family). We assume that the excess is due to the
radioactive contamination of the $^{106}$CdWO$_4$ detector and due
to the HPGe detector passive shield modification (some part of the
passive shield was removed to install the $^{106}$CdWO$_4$
detector). We have also observed gamma quanta with energy 1809
keV, which we ascribe to cosmogenic $^{26}$Al in the aluminium
well of the cryostat. Besides, there is a clear peak at the energy
$262.5(5)$ keV with the counting rate 0.381(4) counts/hour (the
peak is shown in the middle part of Fig. \ref{fig07}). We ascribe
the peak to gamma quanta of the isomeric transition of
$^{113}$Cd$^m$ with energy 263.7(3) keV \cite{Blachot:2010}.

\nopagebreak

\subsection{Limits on $2\beta$ processes in $^{106}$Cd}

\begin{figure}[htb]
\begin{center}
\mbox{\epsfig{figure=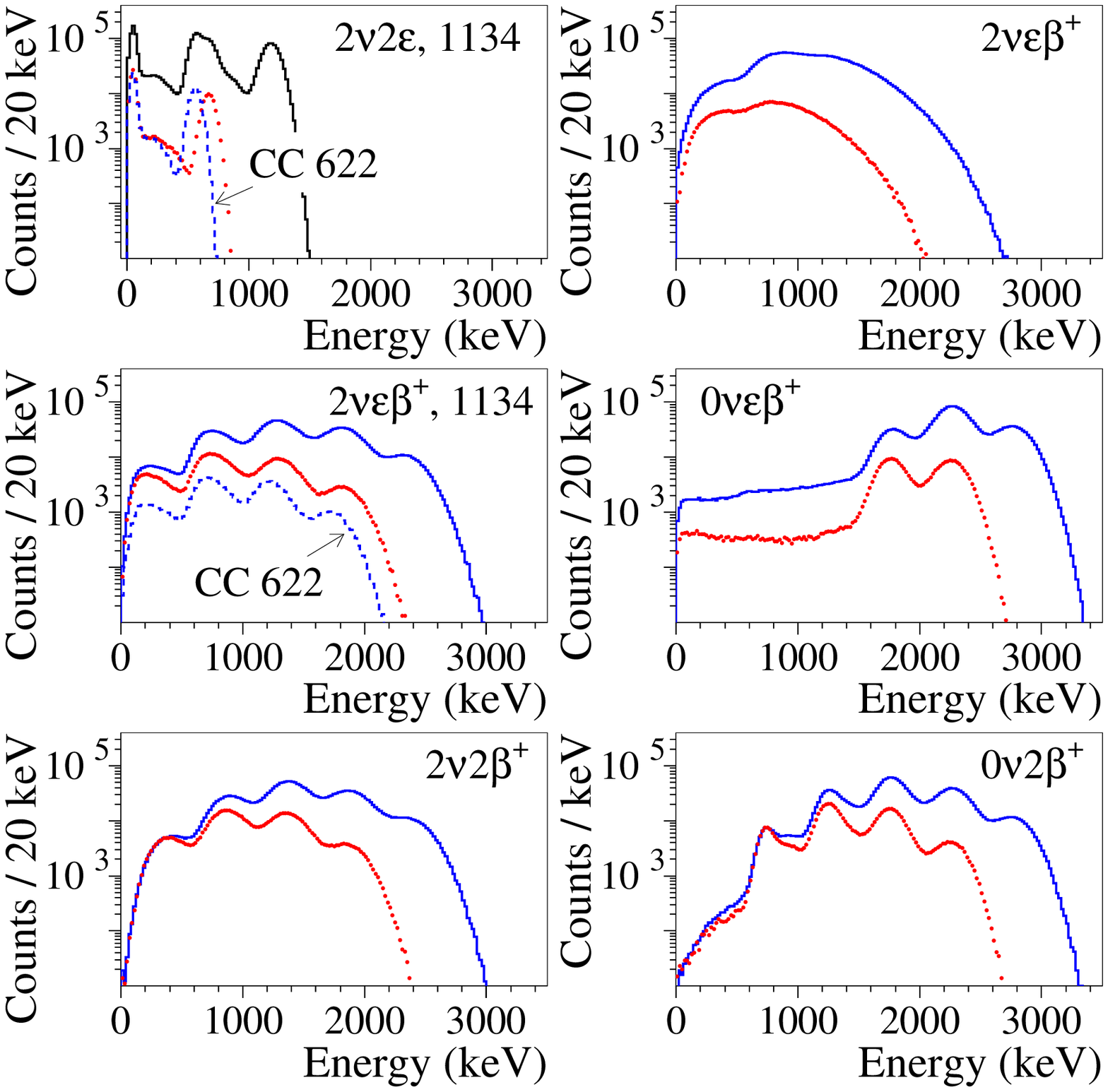,height=10.0cm}} \caption{(Color
online) Simulated response functions of the $^{106}$CdWO$_4$
detector to $2\varepsilon$, $\varepsilon\beta^+$, and $2\beta^+$
processes in $^{106}$Cd without coincidence (solid histograms) and
in coincidence with 511 annihilation $\gamma$ quanta in the HPGe
detector (dotted histograms). Also the response functions of the
$^{106}$CdWO$_4$ detector to  the $2\nu2\varepsilon$ and
$2\nu\varepsilon\beta^+$ decays to the 1134 keV excited level of
$^{106}$Pd  in coincidence with 622 keV gamma quanta  in the HPGe
detector are shown by dashed histograms.} \label{fig08}
\end{center}
\end{figure}

The response functions of the $^{106}$CdWO$_4$ and HPGe detectors
to the $2\beta$ processes in $^{106}$Cd were simulated with the
help of the EGS4 code. The simulated energy distributions in the
$^{106}$CdWO$_4$ detector without coincidence and in coincidence
with 511 keV $\gamma$ quanta (in coincidence with 622 keV gamma
quanta for the $2\nu2\varepsilon$ and $2\nu\varepsilon\beta^+$
decays to the 1134 keV excited level of $^{106}$Pd) in the HPGe
detector are presented in Fig. \ref{fig08}.

There are no peculiarities in the data accumulated with the
$^{106}$CdWO$_4$ and HPGe detectors that could be ascribed to the
$2\beta$ processes in $^{106}$Cd. Therefore only lower half-life
limits can be set by using the formula:

\begin{equation}
\lim T_{1/2} = \ln 2 \cdot N \cdot \eta \cdot t / \lim S,
\end{equation}

\noindent where $N$ is the number of $^{106}$Cd nuclei in the
$^{106}$CdWO$_4$ crystal ($N = 2.42 \times 10^{23}$), $\eta$ is
the detection efficiency, $t$ is the time of measurements, and
$\lim S$ is the number of events of the effect searched for, which
can be excluded at a given confidence level (CL). All the limits
are presented in this paper with 90\% CL.

We have analyzed different data to estimate limits on the $2\beta$
processes in $^{106}$Cd. For instance, to derive the value of
$\lim S$ for the $0\nu\varepsilon\beta^+$ decay of $^{106}$Cd to
the ground state of $^{106}$Pd, the  $^{106}$CdWO$_4$
anticoincidence spectrum was fitted by the model built from the
components of the background and the effect searched for. The best
fit, achieved in the energy interval $1000-3200$ keV, gives the
area of the effect $S=27\pm 49$ counts, thus providing no evidence
for the effect. In accordance with the Feldman-Cousins procedure
\cite{Feldman:1998}, this corresponds to $\lim S=107$ counts.
Taking into account the detection efficiency within the interval
given by the Monte Carlo simulation (69.3\%) and the 95.5\%
efficiency of the pulse-shape discrimination to select $\gamma$
and $\beta$ events, we got the half-life limit: $T_{1/2}\geq
1.5\times 10^{21}$ yr. The excluded distribution of the
$0\nu\varepsilon\beta^+$ decay of $^{106}$Cd to the ground state
of $^{106}$Pd is shown in Fig. \ref{fig06}. The limit is lower
than that obtained in the previous stage of the experiment
\cite{Belli:2012} due to slightly higher background in the high
energy part and worse energy resolution of the $^{106}$CdWO$_4$
detector.

\begin{figure}[htb]
\begin{center}
\mbox{\epsfig{figure=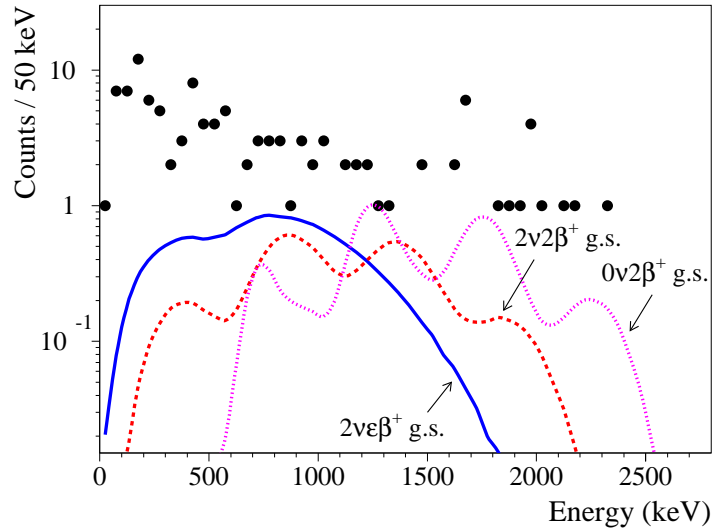,height=7.0cm}} \caption{(Color
online) Energy spectrum of the $^{106}$CdWO$_4$ detector in
coincidence with 511 keV annihilation $\gamma$ quanta in at least
one of the HPGe detectors (filled circles) acquired over 13085 h.
The excluded distributions of different $2\beta$ processes in
$^{106}$Cd are shown by different lines.} \label{fig09}
\end{center}
\end{figure}

The counting rate of the $^{106}$CdWO$_4$ detector is
substantially suppressed in coincidence with the energy 511 keV in
the HPGe detectors. The coincidence energy spectrum of the
$^{106}$CdWO$_4$ detector is presented in Fig. \ref{fig09}. There
are only 115 events in the energy interval $0.05-4$ MeV, while the
simulated background model (built by using the parameters of the
anticoincidence spectrum fit) contains 108 counts. We have
estimated values of $\lim S$ for the $2\beta$ processes in
$^{106}$Cd in different energy intervals. For instance, there are
62 counts in the energy interval $250-1600$ keV, while the
background model contains 68 counts. According to
\cite{Feldman:1998}, one should take $\lim S=8.8$ counts. Taking
into account the detection efficiency of the set-up to the
$2\nu\varepsilon\beta^+$ decay (7.59\%), the part of the energy
spectrum in the energy interval (92.3\%), the selection efficiency
of the pulse-shape discrimination, time and energy cuts used to
build the coincidence spectrum (98.4\% in total), we have obtained
the limit on the $2\nu\varepsilon\beta^+$ decay:
$T_{1/2}^{2\nu\varepsilon\beta^+}\geq 2.0\times 10^{21}$~yr.
However, this limit depends on the model of background which has
some uncertainty related with our limited knowledge of the
radioactive impurities composition and localization in the set-up.
In addition, in some energy intervals (as in the considered above
$250-1600$ keV) the measured number of events is less than in the
model of background. The authors of work \cite{Feldman:1998} suggest
in such a situation to report the so called "sensitivity" of the
experiment defined as the average upper limit that would be
obtained by an ensemble of experiments with the expected
background and no true signal. Following these suggestions we took
a more conservative, background model independent limit on the
number of events of the effect searched for as $\lim S=15.3$
counts, which leads to the following half-life limit on the
$2\nu\varepsilon\beta^+$ decay of $^{106}$Cd to the ground state
of $^{106}$Pd which we accept as a final result:

\begin{center}
$T_{1/2}^{2\nu\varepsilon\beta^+}\geq 1.1\times 10^{21}$~yr.
 \end{center}

\noindent The excluded distribution of the
$2\nu\varepsilon\beta^+$ decay of $^{106}$Pd to the ground state
of $^{106}$Pd is presented in Fig. \ref{fig09} together with the
excluded spectra of the neutrinoless and two neutrino double
positron decay of $^{106}$Cd. The limits on the processes with
positron(s) emission for transitions to the ground and excited
states of $^{106}$Pd were obtained in a similar way mainly by
using the coincidence of the $^{106}$CdWO$_4$ detector with
annihilation $\gamma$ quanta in the HPGe detectors (the limits are
presented in Table~1). A rather high sensitivity was achieved for
the $\varepsilon\beta^+$ and $2\beta^+$ decay channels thanks to
the comparatively high probability to detect at least one
annihilation gamma quantum by the HPGe counters (e.g., 13.6\% for
the $2\beta^+$ decay).

\nopagebreak

\begin{figure}[htb]
\begin{center}
\mbox{\epsfig{figure=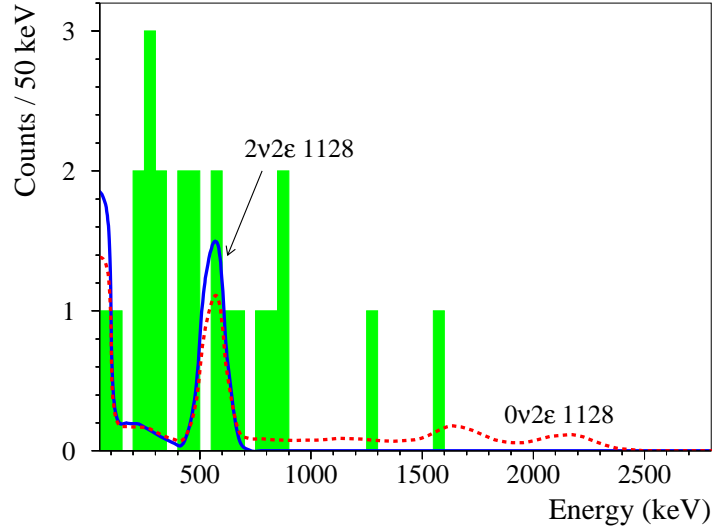,height=7.0cm}} \caption{(Color
online) Energy spectrum of the $^{106}$CdWO$_4$ detector in
coincidence with 616 keV $\gamma$ quanta in the HPGe detector
(filled histogram) acquired over 13085 h. The excluded
distributions of the $2\nu2\varepsilon$ and $0\nu2\varepsilon$
decay of $^{106}$Cd to the 1128 keV excited level of $^{106}$Pd
with detection of 616 keV $\gamma$ quanta in the HPGe detectors
are shown.} \label{fig10}
\end{center}
\end{figure}

Further suppression of the background was achieved by selection of
events in the $^{106}$CdWO$_4$ detector in coincidence with the
intense $\gamma$ quanta in the HPGe counters expected in the
double electron capture in $^{106}$Cd to the excited levels of
$^{106}$Pd. For instance, gamma quanta with energy 616 keV
expected in the $2\varepsilon$ decay of $^{106}$Cd to the excited
level 1128 keV of $^{106}$Pd can be detected by the HPGe counters
with a detection efficiency 2.42\% (2.34\%) in the two neutrino
(neutrinoless) process. As one can see in Fig. \ref{fig10}, there
are only 17 (23) counts in the energy interval $50-700$ keV
($50-2500$ keV). According to \cite{Feldman:1998} one should take
a "sensitivity" estimation of $\lim S=8.4 (9.7)$ counts. The
energy intervals contain 78.2\% (86.2\%) of the simulated
$2\nu2\varepsilon$ ($0\nu2\varepsilon$) spectra. Taking into
account the efficiency of the pulse-shape discrimination, the time
and energy coincidence with the gamma quanta 616 keV (97.3\% in
total), the following half-life limit on the $2\nu2\varepsilon$
($0\nu2\varepsilon$) decay of $^{106}$Cd to the excited level 1128
keV of $^{106}$Pd can be obtained: $T_{1/2}^{2\nu2\varepsilon}\geq
5.5\times 10^{20}$~yr ($T_{1/2}^{0\nu2\varepsilon}\geq 5.1\times
10^{20}$~yr). The excluded distributions of the $2\nu2\varepsilon$
and $0\nu2\varepsilon$ decays of $^{106}$Cd to the 1128 keV
excited level of $^{106}$Pd are presented in Fig. \ref{fig10}.

Most of the limits on the double electron capture in $^{106}$Cd to
the excited levels of $^{106}$Pd were obtained in a similar way
(including the transitions to the levels of $^{106}$Pd with
energies 2718 keV and 2748 keV where the resonant processes are
possible). Typical detection efficiencies for the $2\varepsilon$
processes in $^{106}$Cd to the excited levels of $^{106}$Pd  are
$1.17\%-4.13\%$ depending on the level and decay mode, while the
number of counts varies from 83 to 0  depending on the
energy in the HPGe detectors chosen to build the coincidence
spectra. For example, 83 coincidence counts were
obtained in the energy interval $50-2600$ keV in CC with energy
512 keV in the HPGe counters, while 0 events were detected in the
energy interval $650-2000$ keV in the coincidence with the energy
1160 keV in the HPGe detectors (resonant $0\nu2K$ capture to the
2718 keV excited level of $^{106}$Pd).

\nopagebreak

\begin{figure}[htb]
\begin{center}
\mbox{\epsfig{figure=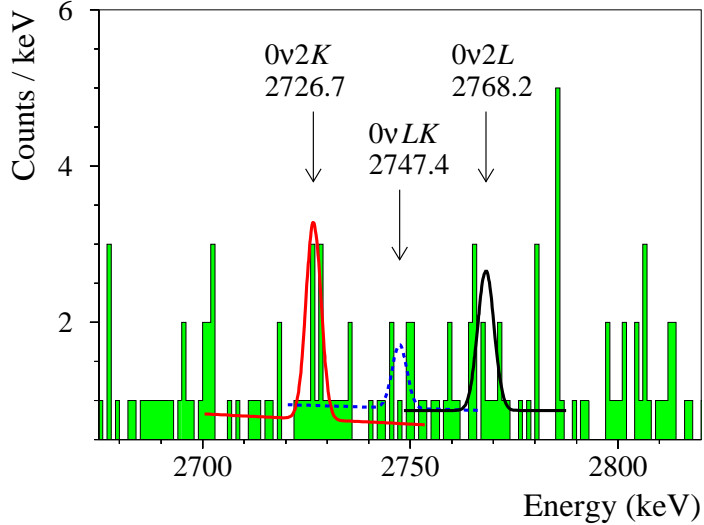,height=7.0cm}} \caption{(Color
online) Part of the energy spectrum accumulated by the HPGe
detector. Excluded peaks expected in the $0\nu 2K$, $0\nu LK$ and
$0\nu 2L$ capture in $^{106}$Cd to the ground state of $^{106}$Pd
are shown. Energies of the expected $\gamma$ peaks are in keV.}
\label{fig11}
\end{center}
\end{figure}

We have also used the data accumulated by the HPGe detectors to
estimate limits on the $2\beta$ processes in $^{106}$Cd. For
instance, in neutrinoless $2\varepsilon$ capture we assume that
the energy excess is taken away by bremsstrahlung $\gamma$ quanta
with energy $E_\gamma = Q_{2\beta} - E_{b1} - E_{b2} - E_{exc}$,
where $E_{bi}$ is the binding energy of $i$-th captured electron
on the atomic shell, and $E_{exc}$ is the energy of the populated
(g.s. or excited) level of $^{106}$Pd. In case of transition to an
excited level, in addition to the initial $\gamma$ quantum, other
$\gamma$'s will be emitted in the nuclear deexcitation process.
For example, to derive a limit on the $0\nu2K$ capture in
$^{106}$Cd to the ground state of $^{106}$Pd the energy spectrum
accumulated with the HPGe detectors was fitted in the energy
interval $2700-2754$ keV by a simple function (first degree
polynomial function to describe background plus Gaussian peak at
the energy 2726.7 keV with the energy resolution FWHM$~=4.4$ keV
to describe the effect searched for). The fit gives an area of the
peak $6.2\pm3.2$ counts, with no evidence for the effect.
According to \cite{Feldman:1998} we took 11.4 events which can be
excluded with 90\% CL. Taking into account the detection
efficiency for gamma quanta with energy 2726.7 keV in the
experimental conditions (1.89\%) we have set the following limit
for the $0\nu2K$ capture of $^{106}$Cd to the ground state of
$^{106}$Pd: $T_{1/2}\geq 4.2\times10^{20}$ yr. Limits on the $0\nu
KL$ and $0\nu 2L$ capture in $^{106}$Cd to the ground state of
$^{106}$Pd ($T_{1/2}\geq 1.3\times10^{21}$ yr and $T_{1/2}\geq
5.4\times10^{20}$ yr, respectively) were derived in the similar
way. The excluded peaks are shown in Fig. \ref{fig11}. We would
like to stress the certain advantages of isotope $^{106}$Cd for
which the high accuracy $Q_{2\beta}$ value is now available.
Thanks to this feature one could distinguish clearly the $0\nu 2K$,
$0\nu KL$ and $0\nu 2L$ modes of the decay using high resolution
HPGe detectors.

\nopagebreak

All the half-life limits obtained in the present work are
summarized in Table~1, where results of the most sensitive
previous experiments are given for comparison.

\nopagebreak
\begin{table}[h!]
\caption{Half-life limits on 2$\beta$ processes in $^{106}$Cd.
The experimental selection (AC anticoincidence, CC in coincidence
at the given energy with HPGe, HPGe using only the HPGe detectors) is also reported.
The results of the most sensitive previous experiments are given for
comparison.}
\begin{center}
\begin{tabular}{|l|l|l|l|l|}
\hline
 Decay                  & Decay     & Level             &  \multicolumn{2}{c|}{ $T_{1/2}$ limit (yr) at 90\% C.L.}\\
\cline{4-5}
 channel                & mode      & of $^{106}$Pd     &  Present work (Data)              & Best previous limit\\
 ~                      & ~         & (keV)             &  ~                                & ~ \\
 \hline
 $2\varepsilon$         & $2\nu $   & g.s.              & $-$                               & $\geq4.2\times10^{20}$  \cite{Rukhadze:2011b} \\
 $2\varepsilon$         & $2\nu $   & 2$^{+}$ 512   & $\geq 9.9\times10^{20}$ (CC 512 keV)  & $\geq1.2\times10^{20}$  \cite{Rukhadze:2011b} \\
 $2\varepsilon$         & $2\nu $   & 2$^{+}$ 1128  & $\geq 5.5\times10^{20}$ (CC 616 keV)  & $\geq 4.1\times10^{20}$ \cite{Belli:2012} \\
 $2\varepsilon$         & $2\nu $   & 0$^{+}$ 1134  & $\geq 1.0\times10^{21}$ (CC 622 keV)  & $\geq 1.7\times10^{20}$ \cite{Belli:2012} \\
 $2\varepsilon$         & $2\nu $   & 2$^{+}$ 1562  & $\geq 7.4\times10^{20}$ (CC 1050 keV) & $\geq 5.1\times10^{19}$ \cite{Belli:2012} \\
 $2\varepsilon$         & $2\nu $   & 0$^{+}$ 1706  & $\geq 7.1\times10^{20}$ (CC 1194 keV) & $\geq 1.1\times10^{20}$ \cite{Belli:2012} \\
 $2\varepsilon$         & $2\nu $   & 0$^{+}$ 2001  & $\geq 9.7\times10^{20}$ (CC 873 keV)  & $\geq 2.9\times10^{20}$ \cite{Belli:2012} \\
 $2\varepsilon$         & $2\nu $   & 0$^{+}$ 2278  & $\geq 1.0\times10^{21}$ (CC 1766 keV) & $\geq 1.6\times10^{20}$ \cite{Belli:2012} \\
 \cline{2-5}
 $2K$                   & $0\nu$    & g.s.          & $\geq 4.2\times10^{20}$ (HPGe)        & $\geq 1.0\times10^{21}$ \cite{Belli:2012} \\
 $LK$                   & $0\nu$    & g.s.          & $\geq 1.3\times10^{21}$ (HPGe)        & $\geq 1.0\times10^{21}$ \cite{Belli:2012} \\
 $2L$                   & $0\nu$    & g.s.          & $\geq 5.4\times10^{20}$ (HPGe)        & $\geq 1.0\times10^{21}$ \cite{Belli:2012} \\
 $2\varepsilon$         & $0\nu$    & 2$^{+}$ 512   & $\geq 3.9\times10^{20}$ (CC 512 keV)  & $\geq 5.1\times10^{20}$ \cite{Belli:2012} \\
 $2\varepsilon$         & $0\nu$    & 2$^{+}$ 1128  & $\geq 5.1\times10^{20}$ (CC 616 keV)  & $\geq 3.1\times10^{20}$ \cite{Belli:2012} \\
 $2\varepsilon$         & $0\nu$    & 0$^{+}$ 1134  & $\geq 1.1\times10^{21}$ (CC 622 keV)  & $\geq 3.5\times10^{20}$ \cite{Belli:2012} \\
 $2\varepsilon$         & $0\nu$    & 2$^{+}$ 1562  & $\geq 7.3\times10^{20}$ (CC 1050 keV) & $\geq 3.5\times10^{20}$ \cite{Belli:2012} \\
 $2\varepsilon$         & $0\nu$    & 0$^{+}$ 1706  & $\geq 1.0\times10^{21}$ (CC 1194 keV) & $\geq 2.5\times10^{20}$ \cite{Belli:2012} \\
 $2\varepsilon$         & $0\nu$    & 0$^{+}$ 2001  & $\geq 1.2\times10^{21}$ (CC 873 keV)  & $\geq 2.3\times10^{20}$ \cite{Belli:2012} \\
 $2\varepsilon$         & $0\nu$    & 0$^{+}$ 2278  & $\geq 8.6\times10^{20}$ (CC 1766 keV) & $\geq 2.1\times10^{20}$ \cite{Belli:2012} \\
 \hline
 Res. $2K$              & $0\nu$    & ~~~~~~~2718   & $\geq 1.1\times10^{21}$ (CC 1160 keV) & $\geq 4.3\times10^{20}$ \cite{Belli:2012} \\
 Res. $KL_{1}$          & $0\nu$    & $4^+$ ~~~2741 & $\geq 8.5\times10^{20}$ (HPGe)        & $\geq 9.5\times10^{20}$ \cite{Belli:2012} \\
 Res. $KL_{3}$          & $0\nu$    & $2,3^-$ 2748  & $\geq 1.4\times10^{21}$ (CC 2236 keV) & $\geq 4.3\times10^{20}$ \cite{Belli:2012} \\
 \hline
 $\varepsilon\beta^+$   & $2\nu$    & g.s.          & $\geq 1.1\times10^{21}$ (CC 511 keV) & $\geq4.1\times10^{20}$  \cite{Belli:1999} \\
 $\varepsilon\beta^+$   & $2\nu$    & 2$^{+}$ 512   & $\geq 1.3\times10^{21}$ (CC 511 keV) & $\geq2.6\times10^{20}$  \cite{Belli:1999} \\
 $\varepsilon\beta^+$   & $2\nu$    & 2$^{+}$ 1128  & $\geq 1.0\times10^{21}$ (CC 511 keV) & $\geq 3.1\times10^{20}$ \cite{Belli:2012} \\
 $\varepsilon\beta^+$   & $2\nu$    & 0$^{+}$ 1134  & $\geq 1.1\times10^{21}$ (CC 622 keV) & $\geq 3.7\times10^{20}$ \cite{Belli:2012}\\
 \cline{2-5}
 $\varepsilon\beta^+$   & $0\nu$    & g.s.          & $\geq 1.5\times10^{21}$ (AC) & $\geq 2.2\times10^{21}$ \cite{Belli:2012} \\
 $\varepsilon\beta^+$   & $0\nu$    & 2$^{+}$ 512   & $\geq 1.9\times10^{21}$ (CC 511 keV) & $\geq 1.3\times10^{21}$ \cite{Belli:2012} \\
 $\varepsilon\beta^+$   & $0\nu$    & 2$^{+}$ 1128  & $\geq 1.3\times10^{21}$ (CC 511 keV) & $\geq 5.7\times10^{20}$ \cite{Belli:2012} \\
 $\varepsilon\beta^+$   & $0\nu$    & 0$^{+}$ 1134  & $\geq 1.9\times10^{21}$ (CC 622 keV) & $\geq 5.0\times10^{20}$ \cite{Belli:2012} \\
 \hline
 $2\beta^+$             & $2\nu$    & g.s.          & $\geq 2.3\times10^{21}$ (CC 511 keV) & $\geq 4.3\times10^{20}$ \cite{Belli:2012} \\
 $2\beta^+$             & $2\nu$    & 2$^{+}$ 512   & $\geq 2.5\times10^{21}$ (CC 511 keV) & $\geq 5.1\times10^{20}$ \cite{Belli:2012} \\
 \cline{2-5}
 $2\beta^+$             & $0\nu$    & g.s.          & $\geq 3.0\times10^{21}$ (CC 511 keV) & $\geq 1.2\times10^{21}$ \cite{Belli:2012} \\
 $2\beta^+$             & $0\nu$    & 2$^{+}$ 512   & $\geq 2.5\times10^{21}$ (CC 511 keV) & $\geq 1.2\times10^{21}$ \cite{Belli:2012} \\
 \hline
\end{tabular}
\label{2b-results}
\end{center}
\end{table}

\subsubsection{Limit on $^{106}$Cd $2\nu\varepsilon\beta^+$ decay matrix element}

An upper limit on the effective nuclear matrix element for the
$2\nu\varepsilon\beta^+$ decay
($M^{2\nu\varepsilon\beta^+}_{eff}$) can be obtained with the help
of the following formula:

\begin{equation}
\lim M^{2\nu\varepsilon\beta^+}_{eff} = 1/\sqrt{\lim
T_{1/2}^{2\nu\varepsilon\beta^+}~G^{2\nu\varepsilon\beta^+}},
\end{equation}

\noindent where $G^{2\nu\varepsilon\beta^+}$ is the phase space
factor for the $2\nu\varepsilon\beta^+$ decay,
$M^{2\nu\varepsilon\beta^+}_{eff} = g_A^2 (m_ec^2)
M^{2\nu\varepsilon\beta^+}$, $g_A$ is the axial vector coupling
constant, and $m_ec^2$ is the electron mass. Using the recently
calculated phase space factors $702\times 10^{-24}$ yr$^{-1}$
\cite{Kotila:2013} and $741\times 10^{-24}$ yr$^{-1}$
\cite{Mirea:2015}, we have obtained the following limit on the
nuclear matrix element for the $2\nu\varepsilon\beta^+$ decay of
$^{106}$Cd to the ground state of $^{106}$Pd:
$M^{2\nu\varepsilon\beta^+}_{eff} \le 1.1$.

\section{CONCLUSIONS}

An experiment to search for $2\beta$ decay of $^{106}$Cd with
enriched $^{106}$CdWO$_4$ crystal scintillator in coincidence with
four crystals HPGe $\gamma$ detector has been completed after
13085 h of data taking. New limits on $2\varepsilon$,
$\varepsilon\beta^+$ and $2\beta^+$ processes in $^{106}$Cd were
set on the level of $T_{1/2}>10^{20}-10^{21}$ yr. The highest
sensitivity was achieved mainly by analysis of coincidence between
the $^{106}$CdWO$_4$ and the HPGe detectors. For the decay
channels with emission of positrons ($\varepsilon\beta^+$ and
$2\beta^+$ decays) a higher sensitivity was achieved in
coincidence with annihilation gamma quanta, while coincidence of
the $^{106}$CdWO$_4$ detector with intensive gamma quanta in the
HPGe counters gives the most stringent limits on the double
electron capture in $^{106}$Cd to excited levels of $^{106}$Pd.
The half-life limit on the two neutrino $\varepsilon\beta^+$
decay, $T_{1/2} \geq 1.1\times10^{21}$ yr, reached the region of
some theoretical predictions
\cite{Hirsh:1994,Barabash:1996a,Toivanen:1997,Rumyantsev:1998,Civitarese:1998,Suhonen:2001}.
By using the half-life limit the nuclear matrix element for the
$2\nu\varepsilon\beta^+$ decay of $^{106}$Cd to the ground state
of $^{106}$Pd can be bounded as $M^{2\nu\varepsilon\beta^+}_{eff}
\leq 1.1$. The resonant neutrinoless double electron captures to
the 2718 keV, 2741 keV and 2748 keV excited states of $^{106}$Pd
are restricted on the level of $T_{1/2} \geq
(8.5\times10^{20}-1.4\times10^{21}$) yr. Unfortunately, the
experiment has no competitive sensitivity to the most probable
$2\nu2K$ channel of the decay due to the high activity of
$^{113}$Cd$^m$ in the $^{106}$CdWO$_4$ crystal scintillator.
Advancement of the experiment is in progress using the
$^{106}$CdWO$_4$ detector in coincidence with two large volume
CdWO$_4$ scintillation detectors in close geometry to improve the
detection efficiency to gamma quanta emitted in the double beta
processes in $^{106}$Cd.

\section{ACKNOWLEDGMENTS}
The authors from the Institute for Nuclear Research (Kyiv,
Ukraine) were supported in part by the project CO-1-2/2015 of the
Program of collaboration with the Joint Institute for Nuclear
Research (Dubna) "Promising basic research on High Energy and
Nuclear Physics" for 2014-2015 of the National Academy of Sciences
of Ukraine. We are grateful to S.R.~Elliott for his note on
presence of $^{116}$Cd $2\nu2\beta^-$ decay in $^{106}$CdWO$_4$
data. We would like to thank Prof. H.~Ejiri for useful
discussions.

\end{document}